\documentclass[useAMS,usenatbib]{mn2e}
\usepackage{graphicx}
\usepackage{float}
\usepackage{color}
\def\apj{ApJ}
\def\aj{AJ}
\def\aap{A\&A}
\def\mnras{MNRAS}
\def\lesssim{\mathrel{\hbox{\rlap{\hbox{\lower3pt\hbox{$\sim$}}}\hbox{\raise2pt\hbox{$<$}}}}}
\def\gtrsim{\mathrel{\hbox{\rlap{\hbox{\lower3pt\hbox{$\sim$}}}\hbox{\raise2pt\hbox{$>$}}}}}

\title[Model-free analysis of quads and substructured galaxies]{Model-free analysis of quadruply imaged gravitationally lensed systems and substructured galaxies}
\author[Woldesenbet and Williams]{Addishiwot G. Woldesenbet$^{1}$\thanks{E-mail: woldesenbet@physics.umn.edu (AGW); llrw@astro.umn.edu (LLRW)} and Liliya L. R. Williams$^{1}$\hfill\\
$^{1}$ School of Physics and Astronomy,
              University of Minnesota,
              116 Church Street SE,
              Minneapolis, MN 55455} 
\begin{document}

\pagerange{\pageref{firstpage}--\pageref{lastpage}} \pubyear{2015}

\maketitle

\label{firstpage}

\begin{abstract}
Multiple image gravitational lens systems, and especially quads are invaluable in determining the amount and distribution of mass in galaxies. This is usually done by mass modeling using parametric or free-form methods. An alternative way of extracting information about lens mass distribution is to use lensing degeneracies and invariants. Where applicable, they allow one to make conclusions about whole classes of lenses without model fitting. Here, we use approximate, but observationally useful invariants formed by the three relative polar angles of quad images around the lens center to show that many smooth elliptical+shear lenses can reproduce the same set of quad image angles within observational error. This result allows us to show in a model-free way what the general class of smooth elliptical+shear lenses looks like in the three dimensional (3D) space of image relative angles, and that this distribution does not match that of the observed quads. We conclude that, even though smooth elliptical+shear lenses can reproduce individual quads, they cannot reproduce the quad population. What is likely needed is substructure, with clump masses larger than those responsible for flux ratio anomalies in quads, or luminous or dark nearby perturber galaxies.
\end{abstract}

\section{Introduction}

The field of multiple image gravitational lensing was born about 35 years ago with the discovery of Q0957+561 \citep{w79} and PG1115+08 \citep{w80}, which were followed by discoveries of many more doubly and quadruply imaged quasars. The theoretical understanding of these systems grew up alongside the growing body of observations. The earliest works found that simple parametric forms for the mass distribution in the lens can account for the systems' observables \citep{young80}, and that image properties, namely positions, shapes, time delays and flux ratios can be derived from the suitably defined lensing potential \citep{bn86,sef92}.

It was also soon realized that strong lensing has degeneracies and invariants, and that these theoretical insights prove useful when lensing is used as a tool. For example, mass sheet degeneracy, and some other degeneracies where different lenses reproduce exactly the same image observables have been well studied \citep{fgs85,gsf88,s00,ld12}, especially in relation to mass modeling, where degeneracies affect the conclusions about the mass profiles of lenses.  The first lensing invariant discovered was the magnification invariant. It states that the sum of signed magnifications for all lensed images of any source within the caustic of a given lens is a constant. This constant can be the same for more than one lensing potential, and is often independent of the parameters characterizing the potential \citep{wm95,d98}. \cite{wm00} and \cite{he01}  showed that image positions can also be used in forming invariants.  Aside from their intrinsic theoretical value, the invariants allow us to rule out some lens models without even fitting the lens. To quote \cite{he01}, ``The major application of lensing invariants is to shortcut the modeling process.''

Because observational uncertainties on the image observables are never zero, not only exact but also approximate degeneracies and invariants are important for the practical work. For example, \cite{ss14} start with an exact invariance transformation that applies to axisymmetric lenses, and show that it survives in an approximate, but still useful form when applied to a wider range of lens models.

Our work in the present paper is in the same vein: we study near invariants that provide useful insights without mass modeling, and have important consequences for mass substructure in lenses. Our analysis is of quads; it involves image positions only, and does not rely on magnifications. 

The relative image locations of a quad are specified by six numbers, say, $(x,y)$ coordinates of any of the three images with respect to the 4th. Lens mass modeling---parametric or free-form---attempts to reproduce all six numbers at once. In \cite{w08} and \cite{us} (hereafter WW12) we started developing a new way of looking at quads, that considers only the three polar angular coordinates of quads, and disregards the three $r$ coordinates.  We showed that information about the mass distribution of certain lens models contained in the azimuthal, or angular coordinates of images around the center of the lens is almost independent of that contained in the radial coordinates, and that the angular coordinates of images show approximate degeneracies. Specifically, if a lens with double mirror symmetry can reproduce the relative images angles of a quad, then all lenses with double mirror symmetry can reproduce the relative image angles of that quad, though for different locations of the (unobservable) source.  Here we continue developing the theory of relative image angles of quads by extending it to lenses represented by purely or approximately elliptical mass profiles with external shear.  The lenses in this general class of mass models are important to study because they fit observed quads well, and are common in parametric lens modeling.

Since we study a wide class of such lenses, and identify similarities, we are able to make conclusions that are independent of a specific lens model. Our analysis suggests that even though these common models are able to reproduce observed quads one at a time, they are unable to reproduce them {\em as a population}. Instead, substructured lenses or luminous or dark nearby perturbers are likely needed.

Because of their importance in the $\Lambda$CDM cosmological model \citep{Moore,k99}, substructure has been the focus of many recent papers. The main method employed for finding substructure uses flux ratios of close pairs or triples of images of quad lenses. In the absence of substructure clumps, i.e. when the potential is smooth, the magnifications of these images obey certain relations \citep{bn86,sw92}.  Substructure induces deviations from these relations, or anomalies \citep{ms98,mz02}. The substructure finding methods based on flux anomalies and image positions are complementary because the former are sensitive to small clumps located close to the images, while the latter are less easily perturbed and respond to larger clumps, $\gtrsim 10^9-10^{10} M_\odot$, located anywhere around the Einstein radius. The two methods also differ in other respects.  Image fluxes are subject to different effects, depending on wavelength, like microlensing by stars in the optical and near infra-red, and propagation effects in the radio, and it is still unclear how major a role these factors play in the observed flux anomalies \citep{xu14}. Image positions are affected only by relatively massive clumps. 

This paper is organized as follows. In Section~\ref{typeII} we define our classification of lens types based on the caustics they produce. In Sections~\ref{powerlaw} and \ref{nonpowerlaw} we discuss the detailed properties of lenses that belong to what we call Type II lenses, which are the focus of this paper.  Readers interested primarily in substructure can skip directly to Section~\ref{realquads}. Section~\ref{conc} summarizes our findings.

\section{Type II lens models}\label{typeII}

Parametric modeling, where parameters such as ellipticity, shear, position angle, etc. identify a lens model, is a commonly used method to theoretically represent gravitational lenses. Simple models obeying certain symmetries are used to fit observed properties of lens systems, such as image positions, time delays, and magnifications. The symmetries could be either in the isopotential or isodensity contours of lenses with different radial density profiles. Examples of such axisymmetric models that are widely used are Elliptic Mass Distribution (EMD) and Elliptic Potential (EP)\citep{KassKov}  with different radial profile such as Isothermal and NFW. The amount of constraining information that can be obtained from a given set of observed images is very limited, and inadequate to fully describe the details of the lensing mass. This means the lens equation is severely under constrained allowing it to admit multiple solutions. But any symmetry in how images are distributed on the lens plane is a direct consequence of the underlying symmetry of the lens and the source position in the source plane. For example, any axially symmetric lens, independent of the radial density profile, would give rise to an Einstein ring for a central source. Therefore categorizing and studying lenses that share sets of symmetries using parameters is a practical approach that does not contradict observational data.

In previous work by the authors (WW12) three classes of lenses were introduced. The classification is based on azimuthal symmetries of isopotential or isodensity contours of lenses, and is independent of the radial density profile. Lens models obeying  twofold symmetry (symmetric about two orthogonal axes) in the lens plane were termed Type I lenses, and breaking this symmetry once resulted in Type II lenses. Type III encompassed all other models.

A more precise categorization, which we will adopt in this paper, can be achieved by using the symmetries of the diamond caustics. Type I lenses are those that give rise to caustics with twofold rotational symmetry\footnote{Objects with n-fold rotational symmetry look the same when rotated by $\frac{360}{n}$ degrees. Twofold symmetry is identical to inversion symmetry through the origin.} and double mirror symmetries, while Type II lenses are the ones that give rise to caustics obeying only twofold rotational symmetry. In this new definition of Type I and II we also limit ourselves to lenses with diamond caustics and exclude those that produce caustics with higher order catastrophes. Such constraint is justified for our analysis which aims to understand the population of observed lens systems. Higher order catastrophes can produce lenses with more than five images, which are not observed in galaxy lenses. Type III lenses include everything except Type I and Type II, for example, substructured lenses. This paper examines Type II models in detail.

Elliptical potentials or mass distributions of arbitrary ellipticity and density profile have twofold and double mirror symmetries. By introducing external shear at a non-zero angle with respect to the ellipticity axes, the double mirror symmetry is broken giving rise to Type II lenses. In this paper we look at Type IIs with power law and NFW density profiles. We identify a lens by three parameters: ellipticity, $\epsilon$, external shear, $\gamma$, and the angle, $\beta$, between shear and the ellipticity principal axis. By definition, $\beta$ lies between $0^{\circ}$ and $90^{\circ}$.

Our basic analysis method uses the relative angular distribution of the four images of quads\footnote{Technically, quads are five image configuration with one of the images invisibly sitting near the very bright center of lenses. For all purposes of this paper, the fifth central image is ignored.} about the center of lens. It is applicable to any lens system, such as quasar-galaxy or galaxy-galaxy, where the center of the lens is known and the four images are point like. This approach was introduced in \cite{w08} and refined in WW12. The method is statistical in nature in the sense that it works with quads as a class to draw conclusions about the mass distribution in galaxy hosts of quad lenses. We will show that properties related to quad image angles are approximately independent of the radial mass density profile of the lens, and that Type I and II lenses cannot account for the observed population of quads.

\section{Power law potentials as examples of Type II}\label{powerlaw}
 
Type II lenses are a wide class. Our strategy is to study a few representative examples of Type IIs in detail, note their similarities and differences, and draw conclusions based on these. Our examples are chosen to be representative of the models used in the lensing literature and to resemble the real galaxy lenses. Furthermore, for computational ease, we use models whose lensing potentials are expressed by simple analytical functions. 

Power law lensing potentials are one example that satisfy all the above criteria. They have a general form,
\begin{equation}
\phi(r,\theta) = b r^\alpha \sqrt{1-\epsilon \cos ( 2 \theta)}+\frac{\gamma}{2} r^2 \cos ( 2[\theta-\beta]).
\label{eq1}
\end{equation}
In the above expression $b$ is the normalization factor, $\epsilon$ is an ellipticity parameter (henceforth ellipticity)\footnote{Note that $\epsilon$ is related to the standard definition of ellipticity \\(1 - axes ratio) as $1-\frac{b}{a} = 1- \sqrt{\frac{1-\epsilon}{1+\epsilon}}$ }, and $r$ is the sky-projected distance from the lens center. The second term is external shear of strength $\gamma$ oriented at an angle of $\beta$ relative to the ellipticity's principal axis. The exponent $\alpha$ in the first term is physically constrained to be between zero ($\alpha<0$ implies decreasing total enclosed mass with increasing $r$), and two ($\alpha>2$ implies that the density is increasing with $r$), but observations constrain it even further to be around $1$ \citep{SLACS3}. Theoretically, we can use any values of $\gamma$ as long as the lens gives rise to quads, whereas $\epsilon$ is limited to be between 0 and 1 by definition. In addition, $\epsilon$ and $\gamma$ are also restricted from above by the requirement that the corresponding isodensity contours are not peanut shaped, which happens, for example, for $\alpha\approx 1.2$ and $\epsilon\approx 0.25$, or $\alpha\approx 1$ and $\epsilon\approx 0.15$.

\subsection {Singular Isothermal Elliptical Potential with External Shear (SIEP+shear)}\label{siep}
  
In this section we use singular isothermal elliptical potential with external shear to demonstrate detailed properties of Type II lenses. SIEP+shear, given by eq.~\ref{eq1} with $\alpha=1$, is chosen because it is the simplest form of power law potential that is semi-analytically treatable.  Different combinations of $\epsilon$, $\gamma$ and $\beta$ give rise to different diamond caustics (based on orientation, elongation of cusps and curving of folds) but all obey the twofold rotational symmetry that identifies Type II lenses. 

\begin{figure}
\includegraphics[scale=.4]{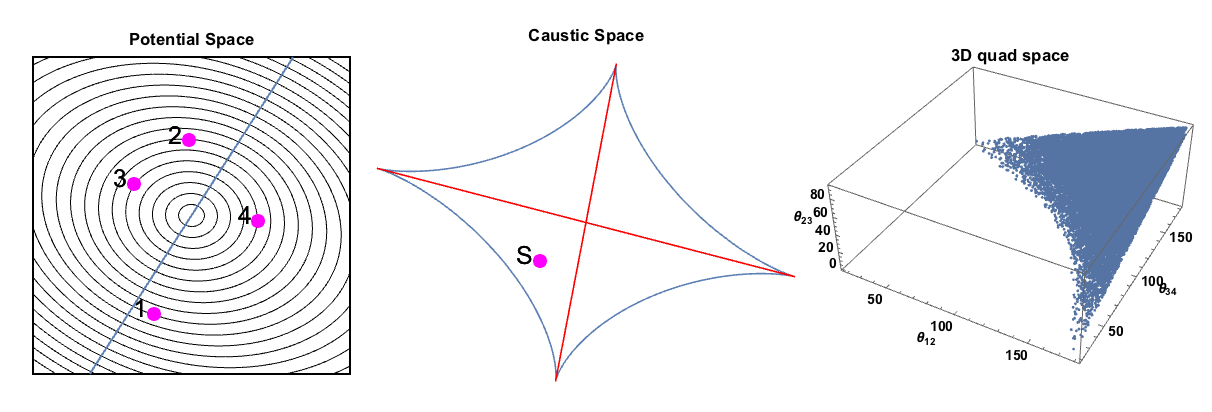}
\caption{Lens Parametrization in three types of spaces:
\textit{Left:} Potential space. A Type II elliptical lens potential contours with external shear oriented along the line (inclined, and going through the center). The four images are numbered based on their arrival time \textit{Middle:} Caustic space, parametrized by the ratio of its diagonals (red lines) and the angle they make at their crossing. \textbf{S} is the source that gives rise to the images shown in the left panel. \textit{Right:} 3D space of quad relative image angles: three dimensional space formed by the three relative angles $\theta_{12}$, $\theta_{34}$ and $\theta_{23}$. Each point in this space represents a quad. The surfaces are generated by images of many randomly distributed sources within the diamond caustic and the cut. }
\label{fig1}
\end{figure}

One interesting property of Type IIs concerns the isopotential contours. The main galaxy, i.e. first term in equation \ref{eq1} with $\alpha=1$ gives rise to contours all of which have the same ellipticity axis, but the addition of external shear results in slightly twisting isopotential contours, i.e. angular orientation of the principal axis of each contour of the total lens potential changes with radius. This is true of the contours in the first panel of Figure~\ref{fig1}, but is hard to see because the degree of twisting is small.

We parametrize a given diamond caustic by angles formed by its diagonals and the ratio of the length of the diagonals. We do not consider the curvatures of the folds, i.e. there could be two diamond caustics of the same diagonal ratios and angles but with different fold curvatures.  In the case of Type I lenses, the caustic diagonals are perpendicular to each other while for Type II caustics, the diagonals form angles other than $90^{\circ}$; see the middle panel of Figure~\ref{fig1}.

Another interesting property of SIEP+shear is that the singularity at the center is not a true critical point; mapping it to the source position using the lens equation gives rise to a set of points forming an oval which acts like an oval caustic, and is called a `cut' \citep{Kov}. Even though this set of points does not satisfy the common definition of a caustic, where the determinant of the Jacobian of the lens mapping is zero, it still acts like one; crossing the cut results in a change of image multiplicity. 

Quads are formed when a source is within the diamond caustic and the cut. The angular distribution of the four images about the center of the lens can be uniquely represented by three relative angles $\theta_{ij}$ (the acute angle between image $i$ and image $j$), where image $i$ is the $i^{th}$ arriving image. Image ordering for synthetic lenses is always known. Images in observed quads can be correctly time ordered in most cases using image morphology \citep{sw03}. 

In this paper we use the same set of relative angles as in WW12: $\theta_{12}$, $\theta_{34},$ and $\theta_{23}$. A 3D space can now be formed using these angles, where a point in the space represents a single quad. Given a lens characterized by ($\epsilon,\gamma,\beta$), one can generate a large number of quads arising from different source positions. Numerical calculation shows that the normalization factor, $b$, has no impact on the angular distribution of the images. This also holds true for the NFW lens discussed later in this paper. The distribution of the corresponding points in the 3D angle space can be used as an alternative way of characterizing the lens. Quads produced by Type I lenses lie on a slightly curved surface, which we called the Fundamental Surface of Quads (FSQ) with a peak at  $(\theta_{12},~\theta_{34},~\theta_{23})=(180^\circ,~180^\circ,~90^\circ)$. Quads from all Type I lenses lie very close to the FSQ, making it a nearly invariant surface, and a useful reference.

On the other hand, quads arising from Type II lenses form two separate surfaces, each usually confined to two different portions in the 3D angle space separated by the FSQ. Based on the positions relative to the FSQ, the two surfaces are identified as upper or lower surface. Like the FSQ, each of these surfaces has two well defined edges which corresponds to $\theta_{12}=180^\circ$ and $\theta_{34}=180^\circ$. These two edges of each of the two surfaces meet to form two peaks, one above and one below the FSQ peak as shown in the left panel of Figure~\ref{fig2}. Each surface sits at the farthest point relative to the FSQ at its peak, i.e. near $\theta_{23}\sim 90^\circ$. The edges of the surfaces come closer to each other at lower values of angle $\theta_{23}$ while their middle parts stay farther apart (right panel of Figure~\ref{fig2}). 
The two $\theta_{23}$ angles of the peaks of the two surfaces of an SIEP+shear lens (or any Type II lens) are supplementary angles (i.e. they add to $180^\circ$). We use the $\theta_{23}$ value of the peak quad of the upper of the two surfaces, $\theta_{23,p}$, as an index to parametrize the surfaces of a given lens in the 3D space.

\subsubsection{Potential, caustic and 3D angles space}\label{genprops}

As described above, Type II lenses can be characterized in three different spaces (Figure~\ref{fig1}). The potential space, where $\epsilon$, $\gamma$ and $\beta$ parametrize the lens, the caustic space where the angles between caustic diagonals and the ratio between their lengths characterize the caustic, and the 3D angles space where $\theta_{23,p}$ value of the peak of the top surface is the characteristic parameter. Now, we describe how the three spaces and their corresponding parameters are related.

{\bf The central source.}~
The source at the center of the lens corresponds to the center of the diamond caustic and is mapped to two pairs of images with each image of a given pair having the same arrival time. We arbitrarily choose one of the first arrivals as image 1 and the other image 2. The same is done with images 3 and 4. The two images with the same arrival time sit at $180^\circ$ of each other on the lens plane, therefore $\theta_{12}=\theta_{34}=180^\circ$. For Type I's, the line connecting opposite images (those with the same arrival time) are orthogonal, so $\theta_{23}=90^\circ$, but this is not the case for Type IIs, where $\theta_{23}$ is either acute or obtuse, depending on the choice of ordering, but the two possibilities are always supplementary angles. The maximum angular separation of images 2 and 3 is attained when the source is at the center of the caustic, and this separation decreases as the source moves radially out from the center. In the 3D angles space, the central source corresponds to two degenerate quads, each located at the peak of each of the two surfaces, at $(180^\circ,180^\circ,\theta_{23,p})$, and $(180^\circ,180^\circ,180^\circ-\theta_{23,p})$.

SIEP+shear has an interesting property connecting the caustic and the images arising from the central source. The two supplementary angles formed by the diagonals of a diamond caustic are same as the $\theta_{23,p}$ angles of the 3D space. The symbolic expressions for the angles formed by the diagonals of the diamond caustics and $\theta_{23,p}$ are rather large when expressed in terms of the parameters $\epsilon$, $\gamma$, and $\beta$. Therefore, we compared them by calculating the difference between them using Wolfram Mathematica for 10,000 different combinations of ($\epsilon$, $\gamma$, and $\beta$) which resulted in exactly zero up to 13 decimal place precision. (In addition, the ratio of the caustic diagonal lengths is the same as the ratio of the radial positions of images 2 and 3 from the center of the lens, but we do not consider the image distance ratios in this paper, so we will not explore this relation any further.)

{\bf Sources on the caustic diagonals.}~
The images of sources on the diagonals of the caustic form the two outer edges of the sheets, ($180^\circ,\theta_{34},\theta_{23}$) and ($\theta_{12},180^\circ,\theta_{23}$). Just like the central source, sources on the diagonals results in degenerate images in terms of arrival time. It is important to note that this degeneracy is only for source positions along the diagonals. All other sources within the caustic are mapped to quads that form the body of the 3D surfaces, with each image arriving at distinctly different time from the rest. Therefore, the two sheets are completely different and not a result of ordering choice of the equally arriving images.

{\bf The rest of the sources.}~
The rest of the two sheets in the 3D angles space comprises quads arising from sources within the body of the diamond caustic.  As shown in the top panels of Figure~\ref{fig3} there is a bifurcation about the diagonal of the caustic. Continuously crossing a diagonal, i.e. moving from one quadrant of the diamond caustics to another (for example, from the black region to the yellow, lighter shade, one) results in a jump between the two sheets of the 3D angles space. Sources on straight lines of constant source position angle in the source plane (various colored lines in the caustic shown in the lower left panel of Figure~\ref{fig3}) result in quads that form non-crossing monotonic, but not necessarily straight curves on the 3D surfaces (lower right panel). 

{\bf The inversion symmetry of Type II lenses.}~
The twofold (inversion) symmetry of the Type II lens diamond caustic also applies to the potential, and is further reflected in the 3D angles space. Source positions from the two opposite quadrants of the caustic generate quads that form one of the sheets. The remaining two quadrants produce the second sheet. In Figure~\ref{fig3} the two yellow, lighter shade, (black) quadrants of the caustic shown in the upper left panel generate the yellow, lighter shade, (black) surface in the 3D angles space shown in the upper right panel. So the inversion symmetry reduces the number of distinct caustic quadrants from four to two. {\em The existence of the two surfaces in the 3D space of relative image angles is the most important characteristic of Type II lenses.}

Note that the properties described in this subsection are not unique to SIEP+shear lenses, but are common to all Type IIs. The inversion symmetry of Type II, which is their defining property, naturally predicts the two surfaces in the 3D angles space. We have tested the existence of the two sheets on many more lenses than are presented in this paper (for example, several Sersic type models), and have confirmed the qualitative behavior of the two sheets. At large $\theta_{23}$ the two surfaces are always found on the opposite sides of the FSQ in the 3D angles space (left panel of Figure~\ref{fig2}), and the separation of the two peaks depends on how much the lens potential deviates from that of Type I.  At small $\theta_{23}$ the edges of the two sheets approach each other and the FSQ (right panel of Figure~\ref{fig2}).

\subsubsection{Near degeneracies of lens models with different sets of ($\epsilon$, $\gamma$, $\beta$)}\label{degen}

As mentioned earlier, we chose to parametrize the distribution of quads in the 3D angles space by $\theta_{23,p}$, the angle between the 2nd and 3rd arriving images of the source located at the center of the lens. The full distribution of quads from a single lens are the two surfaces or sheets (right panel of Figure~\ref{fig1}, or Figure~\ref{fig2}). One may ask if it is adequate to represent the surfaces with just a single point, $\theta_{23,p}$?

Figure~\ref{fig4} shows contour surfaces of constant $\theta_{23,p}$ in $\gamma$ vs. $\epsilon$ vs. $\beta$ space. Any two points on a given surface represent two different lenses in the potential space, each characterized by a different set of ($\epsilon,~\gamma,~\beta$), but sharing the same $\theta_{23,p}$. The two contour surfaces shown have $\theta_{23,p}=91^\circ$ and $\theta_{23,p}=92^\circ$.  In the caustic space these lenses have some commonalities: the diagonals of the caustics intersect at the same angle, since, as previously discussed, these angles are the same as $\theta_{23,p}$. However the folds and cusps of these caustics look different. 

Having matched different lenses based on their $\theta_{23,p}$ value, we would like to see how the corresponding surfaces in the 3D angle space compare. We carry out the comparison in two steps: first we compare the edges and then the bodies of the surfaces. To carry out these comparisons we picked four different lens models, each belonging to the same $\theta_{23,p} = 92^\circ$ contour. We identify each lens as L($\epsilon,~\gamma,~\beta,~\theta_{23,p}$). Table \ref{table1} summarizes these lenses. 

For the first comparison, the ($180^\circ,\theta_{34},\theta_{23}$) edge of the lower of the two 3D surfaces for lenses A-D are plotted in Figure~\ref{fig5}. Visually they seem to form the same line.  But in order to quantify any difference between the edges, we compare fit equations of the lines. The fit equations for Lenses A-D are 
\begin{equation}
\begin{array}{l}
\theta_{23}=0.48617\theta_{34}+0.25271\\
\theta_{23}=0.48617\theta_{34}+0.252017\\
\theta_{23}=0.48621\theta_{34}+0.247357\\
\theta_{23}=0.48627\theta_{34}+0.240303
\end{array}
\end{equation}
respectively, and the angles are expressed in degrees. The average of  the median observational error of the three relative angles is $\sim 0.8^\circ$ (WW12) which is at least two orders of magnitude greater than the difference between the intercepts of the above equations. Similarly, the median observational error of the slope $\theta_{23}/\theta_{34}$ is $\sim 0.006$ which is again two orders of magnitude greater than the difference between the slopes of the four equations above. A corresponding comparison for the other edge of the 3D surface, the ($\theta_{12},180^\circ,\theta_{23}$), yields similar results. Therefore, the difference between the edges of these four, and presumably other different SIEP+shear lenses in the 3D space is negligible for all practical and observational purposes. 

For the second comparison we generate two sets of random source positions on one of the two surfaces of each lens. The quads from the first set are used to obtain an interpolation fit equation, which is then compared to the quads of the second set. This is our calibration, or cross-validation procedure. The comparison is done with the root mean square of errors, RMSE. The cross-validation for Lens A yields RMSE of $0.005^\circ$. The RMSE's of Lenses B-D, when compared to the fit equation of Lens A are $0.011^\circ$, $0.029^\circ$, and $0.017^\circ$, respectively (see Table 1). A visual indication of how close these surfaces look is provided by Figure~\ref{fig10}, which plots a certain projection of the 3D angles space (explained in detail in Section~\ref{realquads}) for Lenses A-D. The largest deviation is for Lens C, represented by (green) diamond symbols, which extend lower along the vertical axis by about $0.1$ degree compared to the other three lenses. In general, lenses with largest deviations from the rest of their family members are those with larger ratios of $\epsilon/\gamma$. 

The RMSE values and Figure~\ref{fig10} show that the surfaces are not mathematically identical, but because the deviations between lenses sharing the same $\theta_{23,p}$ are smaller than the typical observational error (even for Lens C), observationally speaking these four SIEP+shear lenses of different lens parameters ($\epsilon,~\gamma,~\beta$) are degenerate in the 3D angles space, i.e. in terms of the relative image angles.  In other words, if a lens model described by ($\epsilon_1,~\gamma_1,~\beta_1$) is found to fit the relative angles of an observed quad, then any other lens characterized by the same $\theta_{23,p}$, but a different set ($\epsilon_2,~\gamma_2,~\beta_2$) will also be able to reproduce the image angles, though for a different location of the source. 
 
So the answer to the question raised at the top of this section is {\em yes}, a single point in the 3D angles space, $\theta_{23,p}$, adequately represents the rest of the surfaces. 

We note an interesting consequence of the degeneracy described above. Taking a $\beta$=const two dimensional slice through the surfaces in Figure~\ref{fig4} we get contours plotted in Figure~\ref{fig6}. These contours are symmetric about $\epsilon=\gamma$ line, which implies that $\epsilon$ and $\gamma$ are interchangeable. This is unexpected because this symmetry is not at all obvious from the equation of the lens potential, eq.~\ref{eq1} with $\alpha=1$.
Later we will see that this symmetry applies to other potentials if the isopotential contours are purely elliptical (as defined in Section \ref{Non-iso}).

Note that in this paper we do not consider image distances from the lens center; two lenses degenerate in image angles need not be degenerate in image distances.

\subsection {Non-isothermal Elliptical Potentials with External Shear}\label{Non-iso}

As discussed when introducing the general power law potential, eq.~\ref{eq1}, $\alpha$ is observationally constrained to be $\sim 1$ \citep{SLACS3}. Therefore, we extend our discussion of Type II lenses only somewhat beyond this value; we use $\alpha=0.9$ and $\alpha=1.2$. We expand $\alpha$ up to 1.2 so that we can explore lenses with higher values of ellipticity. Going beyond these limits results in the mass density contours becoming peanut shaped even for moderate values of ellipticity $\epsilon$. For $\alpha$ = 1.2 the two sheets in the 3D space starts crossing for small values of $\theta_{23}$ when the ratio $\epsilon/\gamma \gtrsim 1$.

We note that the way $\alpha$ is introduced in eq.~\ref{eq1} creates lenses whose isopotential contours due to the main galaxy are not exactly elliptical; we call these hybrid potentials. The pure elliptical potentials are obtained by raising what we call the elliptical radius, $r\sqrt{1-\epsilon \cos{2 \theta}}$, to the power of $\alpha$ instead of raising just $r$. But the angular distribution of quads generated by pure elliptical potentials are not distinctly different from that of SIEP+shear, therefore our discussion in this section focuses on hybrid potentials.  

The $\gamma$ vs. $\epsilon$ contour is not symmetric about $\gamma = \epsilon$ line, with the contours rotated clockwise for $\alpha= 0.9$ and counter clockwise for $\alpha= 1.2$; see Figure~\ref{fig5}. But for pure elliptical lenses defined above, the contours remain symmetric independent of $\alpha$ and this symmetry extends to NFW profiles discussed in the next section, as already alluded to at the end of Section~\ref{degen}.

We would like to know if the degeneracy property of SIEP+shear (Section~\ref{degen}) applies to non-isothermal lenses, i.e. if these lenses can be adequately characterized by just $\theta_{23,p}$.  For that purpose we generate quad distributions for Lenses F - I; see Table~\ref{table1}. Lens F and G have $\alpha=1.2$, while H and I have $\alpha=0.9$ and all share the same peak of $\theta_{23,p}=93^\circ$. We use the second of the two tests introduced in Section~\ref{degen}. We find fit equations for one of the two surfaces of Lenses F and H, and compared these to the actual surfaces of Lenses G and I.

The cross-validation RMSE for Lenses F and H are $0.013^\circ$ and $0.003^\circ$. The RMSE for Lens I as predicted by H is $0.264^\circ$. The RMSE for Lens G as predicted by F is $0.551^\circ$. However, the high $\epsilon$ to $\gamma$ ratio of Lens F ($\epsilon/\gamma=4.5$) results in its two surfaces in 3D angles space crossing each other, which makes it hard to separate out a single surface and find a fit equation for it.  Therefore the latter value can be compared only approximately to other RMSE values.

This comparison of different lenses within the same family (F vs. G and H vs. I) shows that non-isothermal hybrid potential lenses sharing the same peak $\theta_{23,p}$ are approximately degenerate in the 3D angles space (all RMSEs quoted above are smaller than observational uncertainty), but are not as similar to each other as those within the SIEP+shear family.

We note, however, that this comparison might be affected by the fact that these calculations were done numerically; SIEP+shear potential is simple enough to be amenable to semi-analytical calculations, but non-isothermal hybrid potentials require fully numerical computations, and there are two places where numerical errors can creep in: (i) when selecting ($\epsilon,~\gamma,~\beta$) sets from the same $\theta_{23,p}$ contour, and (ii) when doing forward lensing of sources to produce images and measure their polar angles. These numerical errors could contribute to the differences in the 3D surfaces. In the next section, where these errors are also an issue, we carry out a test to assess the error arising from (i).

As with isothermal lenses, the models that differ the most from the rest of the family are those with larger $\epsilon/\gamma$ ratios. For $\alpha=1.2$ larger $\epsilon/\gamma$ ratio also make the two surfaces in the 3D angle space cross the FSQ and each other. Nevertheless, the differences quoted above are at most of the order of observational errors. Therefore observations based on the relative image angles of quads can not discriminate between any of the lenses within the family given by eq.~\ref{eq1}, as long as $\epsilon/\gamma\lesssim 2-2.5.$

\section{Non Power law potentials as examples of Type II}\label{nonpowerlaw}

In this section we broaden our exploration of Type II lenses to include models with non power law density profiles. We chose NFW radial density profile \citep{NFW}, which has an analytical form for its lensing potential \citep{NFWPotential}. Though the central density profiles of lensing galaxies are significantly affected by baryons and so typically have steeper than NFW slopes, we use this profile to explore how non power law profiles behave as Type II lenses. In order to make a Type II lens with NFW radial density profile we introduce external shear and ellipticity to the potential, 
\begin{equation}
\phi(r,\theta) = b\Biggl (\ln^2{\frac{r}{2}}-\frac{1}{4} \ln^2{\frac{1+\sqrt{1-r^2}}{1-\sqrt{1-r^2}}}\Biggr) +\frac{\gamma}{2} r^2 \cos ( 2[\theta-\beta])
\label{eqNFW}
\end{equation}
where $r={x \sqrt{1-\epsilon \cos(2 \theta)}}/{r_{s}}$, $x$ is the polar radius in the lens plane, $r_{s}$ is the scale factor of NFW density profile, and $b$ is a normalization that is a function of $r_{s}$ and the characteristic density. The rest of the variables are as defined in previous sections. This potential is defined for $r<1$, i.e for $x<r_{s}$.

The diamond caustic of this potential has the symmetry of Type II lenses, and gives rise to two surfaces in the 3D space of relative image angles of quads. As shown in upper right panel of Figure~\ref{fig6}, $\gamma$ vs. $\epsilon$ contour map of constant $\theta_{23,p}$ is symmetric with respect to $\gamma=\epsilon$ line. Lenses J and K in Table~\ref{table1} are two different lenses in potential space whose two surfaces each have the same peak, $\theta_{23,p}$, in the 3D angles space. To check if these surfaces of two different lenses are degenerate in 3D angles space, the second test of Section~\ref{degen} is implemented. A cross-validation RMSE for Lens J is $0.006^\circ$ while the RMSE of Lens K as predicted by Lens J is $0.181^\circ$. 

Again, numerical errors could contribute to the difference between the 3D angles space surfaces of these two lenses. To show that this is likely to be the case we made use of the symmetry with respect to the $\gamma=\epsilon$ line mentioned in the previous paragraph. We generated 3D angle space surfaces for sets of lenses related by this symmetry: Lenses K and KK, and J and JJ in Table 1 are two examples. All four lenses share the same $\theta_{23,p}=93^\circ$, but the $\epsilon$ and $\gamma$ parameters of J and K were picked from numerically generated output, whereas those of K and KK  (J and JJ) were obtained by swapping $\epsilon$ and $\gamma$ values, which yields exact $\epsilon$ and $\gamma$ parameters. The RMSE from the comparison of Lenses J and JJ (K and KK) yields $0.007^\circ$ ($0.036^\circ$), values which are considerably smaller that $0.181^\circ$ quoted in the previous paragraph.

Whatever the source of the discrepancy between lenses such as J and K, the difference is still less than the observational errors, so elliptical NFW+shear lenses with the same $\theta_{23,p}$ are nearly degenerate in terms of image relative angles.

\section{Implications for the Observed Quads and Substructure in Lens Galaxies}\label{realquads}

In this section we rely on the basic property of Type II lenses, namely that the quads of each lens generate two distinct surfaces in the 3D space of image relative angles, such as shown in Figure~\ref{fig2} (Section~\ref{genprops}), and that the two sheets are qualitatively similar for all Type II lenses that are suitable as models for observed quads. 

\parskip-0.1in
\subsection{Why degeneracies are useful}
\parskip0in
To draw conclusions about the broad class of lenses used as models
one needs to explore a wide range of Type II lens models that are used in the literature to model quads. This task is made easier by the results of Sections~\ref{powerlaw}-\ref{nonpowerlaw}, where we explored detailed properties of three different representative classes of parametric models as examples of Type II lenses, and showed that lenses within the same class, like SIEP+shear, or elliptical NFW+shear exhibit approximate degeneracies, in the sense that lenses with the same angle $\theta_{23,p}$ have very similar distribution of quad relative angles in the 3D angle space. As mentioned earlier, we also verified that other commonly used profiles, most notably Sersic, also share these properties.

Furthermore, we now show that this degeneracy extends to lenses with different radial density profiles sharing the same peak, $\theta_{23,p}$, as long as the ratio $\epsilon/\gamma$ is smaller than $2-2.5$. To show this we introduce another lens, Lens E (SIEP+shear), which has the same $\theta_{23,p}$ as Lenses F - K. Interpolation fit for Lens E gives a cross-validation RMSE=0.003 degrees. Then, fitting Lenses G (hybrid with $\alpha =1.2$), H (hybrid with $\alpha =0.9$), and J (NFW) give RMSE of 0.146, 0.011, and 0.085 degrees, respectively (Table 1), which are all smaller or of the order of observational uncertainty. This is an important result: most Type II lenses which have the same angle between 2nd and 3rd arriving images of the central source are nearly degenerate in terms of their image angles, for all sources.

The existence of this degeneracy, even if approximate, reduces the number of lens models one has to consider. It follows that we have to consider only a set of lenses with different $\theta_{23,p}$, and any density profile (power law or curved in log-log space, like NFW, with the projected density slope not too different from 1 and any set of ($\epsilon,~\gamma,~\beta$), as long as $\epsilon/\gamma\lesssim 2-2.5$. The restrictions on $\alpha$ and $\epsilon$ are because the degeneracies break down outside of the specified ranges. However, almost all of the lens models used to fit observed quad systems belong to the set of degenerate lenses. It is also important to stress that the general shape of the surfaces formed by quads in the 3D angles space is the same for all Type II lenses, even if they are not degenerate.

\parskip-0.1in
\subsection{Type II lenses cannot fit the population of observed quads}
\parskip0in
Now we are ready to compare Type II lenses with the observed quad population. We will work in two dimensions instead of the three dimensions of the 3D angles space. We use the FSQ as a reference surface, and plot the vertical ``distance'' between a quad in the 3D relative angles space and the FSQ, called $\Delta\theta_{23}$, versus $\theta_{23}$. This plane is shown in Figure~\ref{fig7}, together with the quads data (black filled symbols with error-bars) originally presented in WW12.  FSQ in this plot is the $\Delta\theta_{23}=0$ line. Using this same diagnostic tool, Type I lenses were shown to be inconsistent with the observed quad population (WW12). Specifically, the deviation of even extreme Type I lens models from the FSQ is not enough to account for the wide spread of observed quads in $\Delta\theta_{23}$ vs. $\theta_{23}$ plane, especially at the lower values of $\theta_{23}$.

Gray regions in Figure~\ref{fig7} represent four examples (with two sheets per lens) of Type II lenses. Larger values of $\gamma$ or $\epsilon$ move the gray surfaces further away from the FQS. Looking at the gray surfaces shown in that Figure, and remembering that many more similar surfaces can be drawn, both closer and further away from the FSQ, one may conclude that the whole wide class of Type II lenses---the most commonly used parametric lens models---should be able to reproduce the quad data. This impression comes about because one can find a Type II model, i.e. a set of ($\epsilon,~\gamma,~\beta$) that would reproduce any individual quad (with the possible exception of the extreme quads at $|\Delta\theta_{23}|\gtrsim 15^\circ$), or in the parlance of Figure~\ref{fig7}, one can find a Type II model whose surfaces go through any given quad. 

However, this impression that Type IIs can reproduce the quad population is false, due to two reasons, with the second one being more important.

(1) In order to reproduce quads at small $\theta_{23}$ Type II lenses would need to have high values of ellipticity and shear. For example, the full SLACS lens samples were fit with SIE mass profiles plus external shear model with median (maximum) values of $\gamma$ and $\epsilon$ of 0.05(0.27) and 0.21 (0.49), respectively \citep{SLACS5}, and the corresponding values from \cite{sluse12} for 14 quads are 0.12 (0.33) and 0.20 (0.49). The higher end of these parameter ranges appear to be larger than what should be expected in a realistic lensing case. Using monte carlo simulation and considering the effects of environment \cite{Wong} have found the typical value of total shear to be 0.08 with highest value of 0.17, and \cite{Dalal1} calculated a typical value for shear of 0.03.\footnote{Note that these ellipticity values refer to mass, so cannot be compared directly to our $\epsilon$.} Similar results were obtained by \cite{k97} and \cite{hs03}. These works imply that Type II lenses used as models for some observed quads have higher $\gamma$ values than realistically expected.


(2) In Figure~\ref{fig7} the relative angles of all quads at $|\Delta\theta_{23}|< 12^\circ$ and $\theta_{23}<45^\circ$ can be modeled with Type II lenses shown in that plot as gray surfaces. However, these same Type II models also predict a large population of quads at $\theta_{23}\gtrsim 50^\circ$ and $|\Delta\theta_{23}|\gtrsim5^\circ$. In fact, any Type II model predicts that about $1/3$ of its quads should have $\theta_{23}\!>\!60^\circ$. This is in contradiction to the observations which show virtually no quads in that region.
We conclude that even though Type II lenses are able to model most quads individually, they introduce a serious problem because they predict the existence of quads at large $\theta_{23}$ and $\Delta\theta_{23}$ of at least a few degrees. 


\parskip-0.1in
\subsection{Is our conclusion affected by a lens selection bias?}
\parskip0in
Since the population of quads we use does not represent a homogeneous sample, one may wonder if selection biases are responsible for the distribution of observed quads in the 3D angle space and hence in the $\Delta\theta_{23}$ vs. $\theta_{23}$ plane. In particular, can selection effects negate the argument in (2) above?

To see that this is not the case let us recall how angle $\theta_{23}$ of a quad is related to the source's location within the caustic. Quads with large $\theta_{23}$ originate from the central regions of caustics, while those with small $\theta_{23}$, from the outer regions, adjacent to the folds and cusps. To account for the distribution of observed quads described in (2) one would need to have the central quads arising from Type II lenses suppressed by some selection bias, while not suppressing the central quads arising from Type I lenses. 
It would be hard for a selection bias to accomplish that. For example, magnification bias would bias would suppress central quads because they have smaller magnifications. However, it could not distinguish between Type I and Type II lenses, which is what would be needed to explain the distribution seen in Figure~\ref{fig7}.


\parskip-0.1in
\subsection{Modeling individual quads vs. modeling quad population}
\parskip0in
We stress that our analysis presented above differs significantly from lens modeling of individual quads. Unlike lens mass modeling, parametric or free-form, which is done with one lens system at a time, we study the population properties of quads, and compare them to the generic properties of Type II lenses.  This is why we refer to our method as model-free analysis of quads.  As pointed out above, and can be seen from lens modeling literature, most individual quads can be successfully modeled with Type II lenses. However, these same lens models also predict the existence of quads which are not observed, and whose absence would be hard to explain by a selection effect. This predicted, but unobserved population of quads is most clearly seen in the space of relative images angles, as is done in Figure~\ref{fig7}.
 
\parskip-0.1in
\subsection{Type III lenses}
 \parskip0in
This leaves us with Type III lenses, for example substructured lenses. A preliminary analysis gives promising results. As depicted in Figure~\ref{fig8}, a better match to the distribution of quads in the $\Delta\theta_{23}$ vs. $\theta_{23}$ plane is achieved with the introduction of substructure in the main lens. By increasing the clumpiness of the lens, the gaps in the peak region, near large $\theta_{23}$, of the two surfaces of quads decreased significantly and looks more like the observational data. However, too much clumpiness could lead to higher order catastrophes in the caustic which would result in observationally unsupported systems with higher image multiplicities. In the examples used in Figure~\ref{fig8} this is not a concern since even the clumpiest of the two substructured lenses (bottom row) produce more than five images for less than $1\%$ of the total sources.

Figure~\ref{fig9} shows that one does not need to use a highly substructured lens to reproduce the population distribution of observed quads. Here we start with a substructure-less Type I lens (top row)  that gives rise to quads which are closely packed on the horizontal axis, i.e. the FSQ, failing to match the spread of observed quads.  The introduction of substructure (bottom left) does not perturb the shape of the outer isodensity contours, but the resulting distribution of quads is dispersed in the $\Delta\theta_{23}$ vs. $\theta_{23}$ plane in the same fashion as the observed quads (bottom right).  

This is just a limited foray into Type III lenses. The conclusion of this Section is that Type II lenses cannot reproduce the distribution of image angles of observed quads, and that possibly substuctured lenses, or additional nearby perturber galaxies are needed. We leave quantitative work on these lenses, and detained comparison with the observed quad distribution for a later paper.

\section{Conclusions}\label{conc}

In this paper we presented a technique to explore the properties of quadruple image gravitational lens systems. In general, the relative distribution of quad's four images about the center of a lens can be fully described in 6D space of three relative angles and three relative radial distances in the lens plane. Our technique uses only the 3D subspace of relative angles which is orthogonal to the remaining 3D subspace of relative distances. 

\begin{figure}
\begin{flushleft}
\includegraphics[scale=.57]{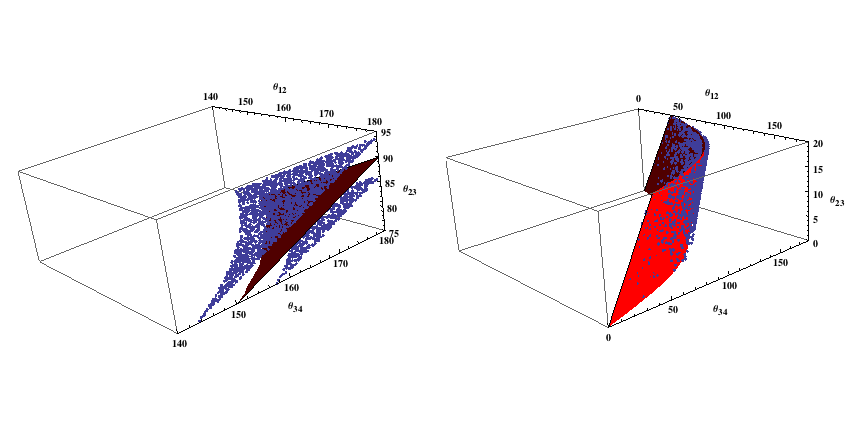}
\caption{3D angles space: Type II lenses produce quads that form two sheets on either side of the FSQ, shown as the smooth surface in both panels. (Note that it looks dark red from the left and bright red from the right). Quads belonging to the Type II lens are represented by light blue dots. \textit{Left}: The top portion, $\theta_{23}>75^\circ$, of the 3D space and the surfaces. The peak of the FSQ is at $\theta_{23,p}=90^\circ$. The peaks of the two surfaces of Type II lens are at $\theta_{23,p}\sim 94^\circ$, and $\theta_{23,p}\sim 86^\circ$. \textit{Right}: The lower portion of the same 3D space, for $\theta_{23}<20^\circ$.}
\label{fig2}
\end{flushleft}
\end{figure}

\begin{figure}
\begin{flushleft}
\includegraphics[scale=.5]{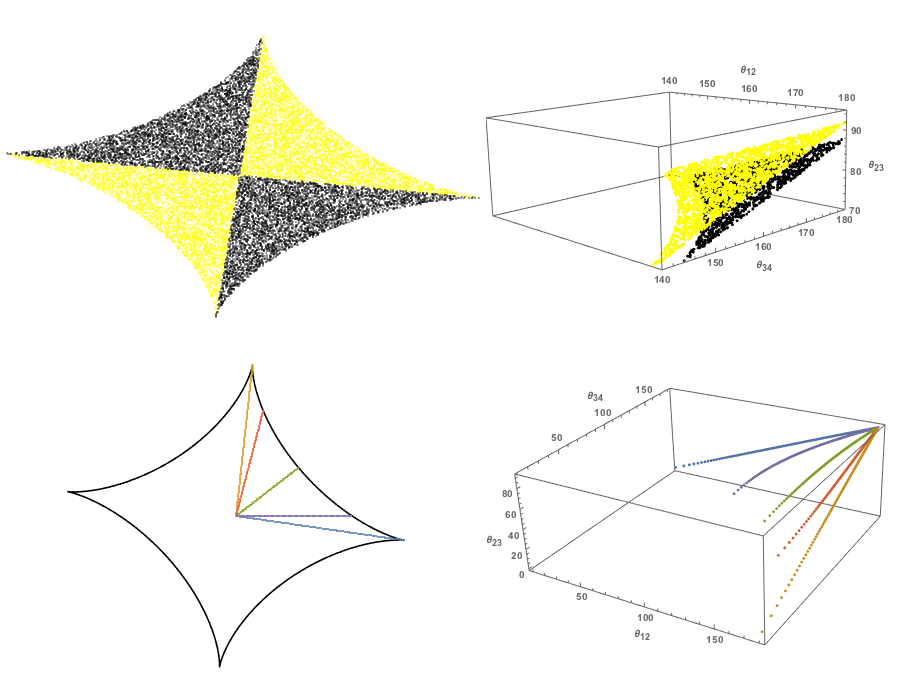}
\caption{Mapping of the caustic space to the 3D space of relative image angles. \textit{Top}: Each pair of opposite quadrants of the diamond caustic (for example, the two black quadrants) map into one of the two surfaces in the 3D space (black). This is a reflection of the twofold symmetry of Type II lenses. \textit{Bottom}: Sources along constant position angle in the caustic plane (various colored lines) are mapped to non-crossing curves in the 3D angles space (similarly colored lines).}
\label{fig3}
\end{flushleft}
\end{figure}

\begin{figure}
\includegraphics[scale=.4]{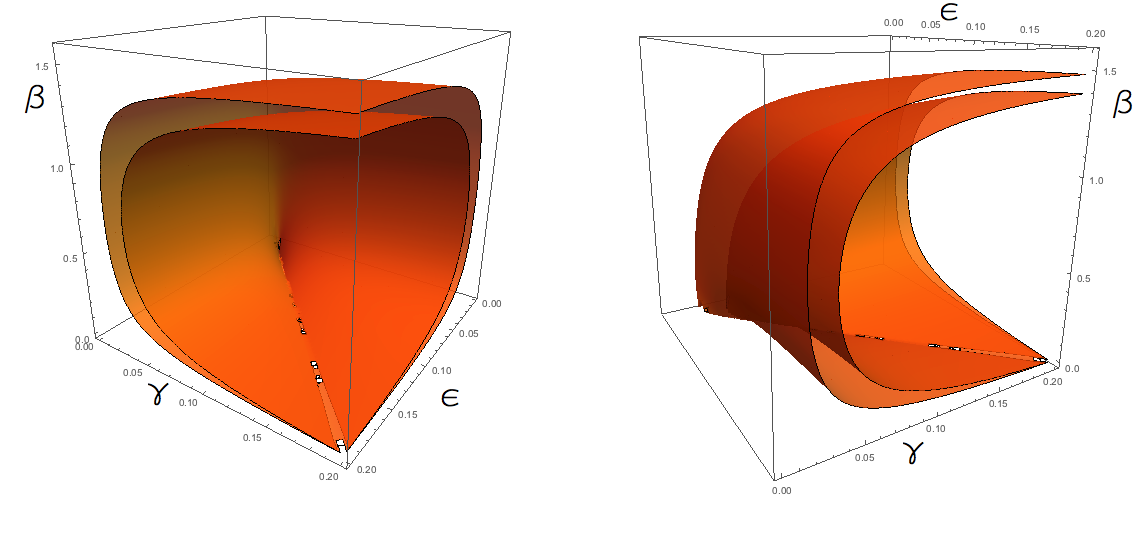}
\caption{Contour surfaces of constant $\theta_{23,p}$ for SIEP+shear lens in $\gamma$ vs. $\epsilon$ vs. $\beta$ space. $\beta$ is in radians. The outer (inner) surface is for $\theta_{23,p}= 91^\circ$ ($\theta_{23,p}=92^\circ$). The jagged line corresponding to $\beta=0$ and $\gamma=\epsilon$ is the result of no quads being formed for those parameters. Two orientations of the same space are shown.}
\label{fig4}
\end{figure}

\begin{figure}
\includegraphics[scale=.45]{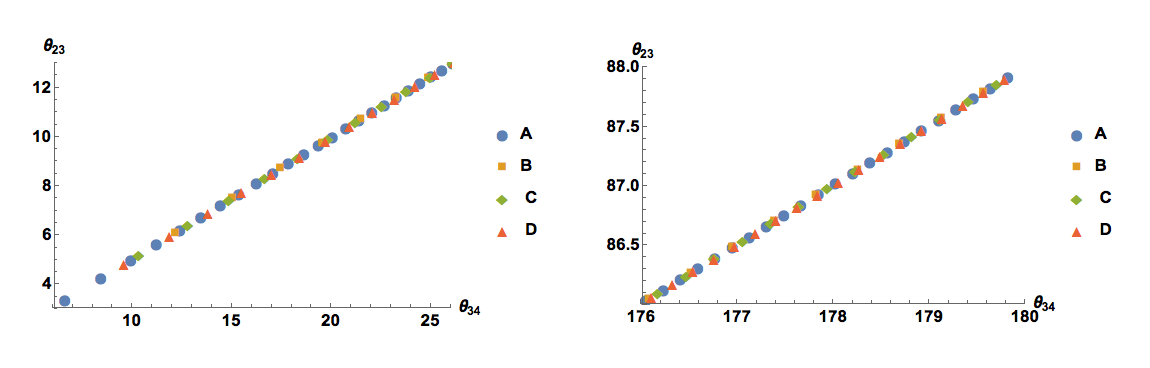}
\caption{The $\theta_{12}=180 ^\circ$ edge of the lower of the two surfaces, in the 3D space of relative angles, for four SIEP+shear lenses sharing the same peak $\theta_{23,p}$. \textit{Left:} Bottom end of the edge. \textit{Right:} Top end of the edge. The parameters of the four lenses A-D are given in Table 1.}
\label{fig5}
\end{figure}

\begin{figure}
\includegraphics[scale=.4]{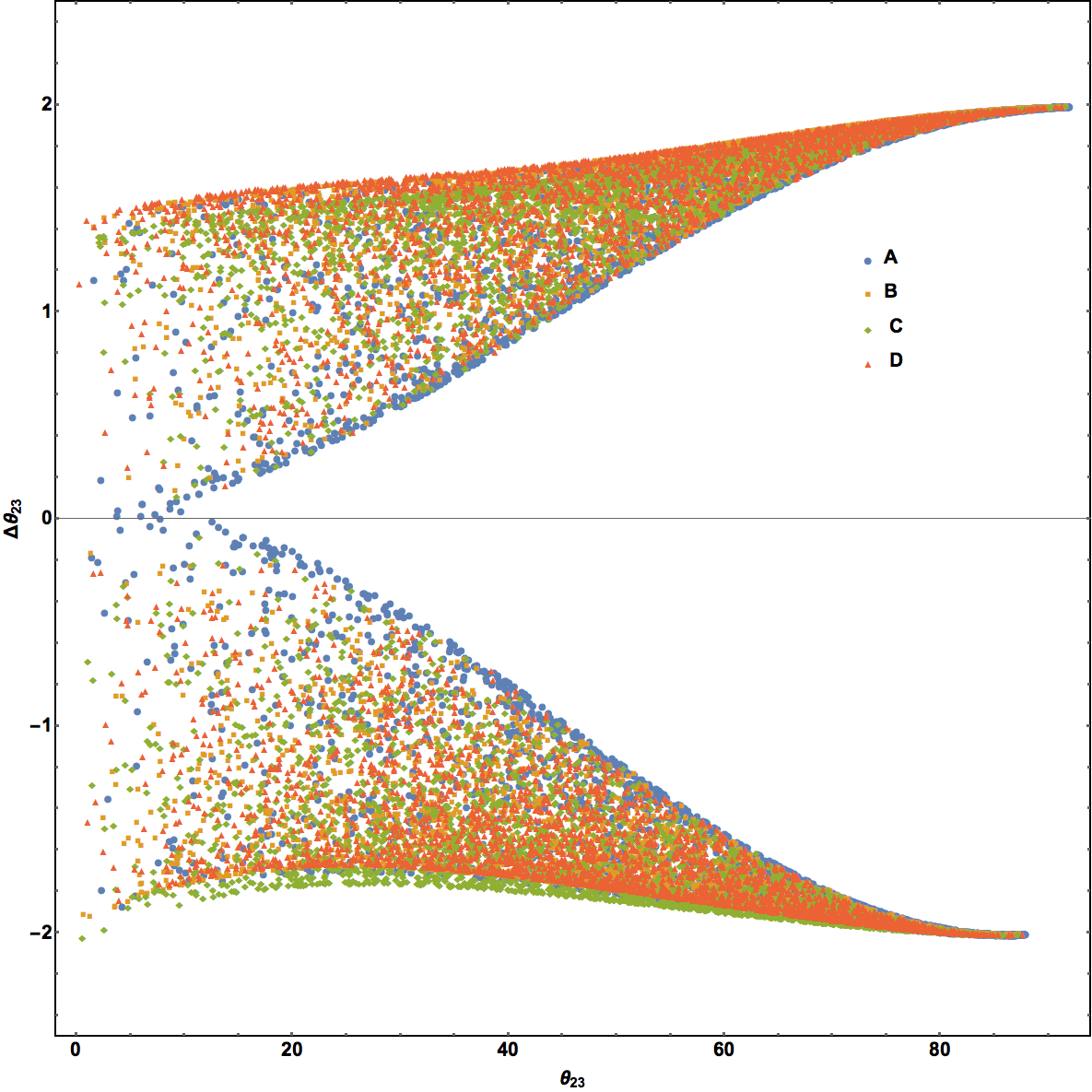}
\caption{Pojection of Lenses A-D with parameters as given in Table \ref{table1} . Lens C, diamond (green), slightly deviate from the other three lenses by saging at the folds.}
\label{fig10}
\end{figure}

\begin{figure}
\begin{flushleft}
\includegraphics[scale=.4]{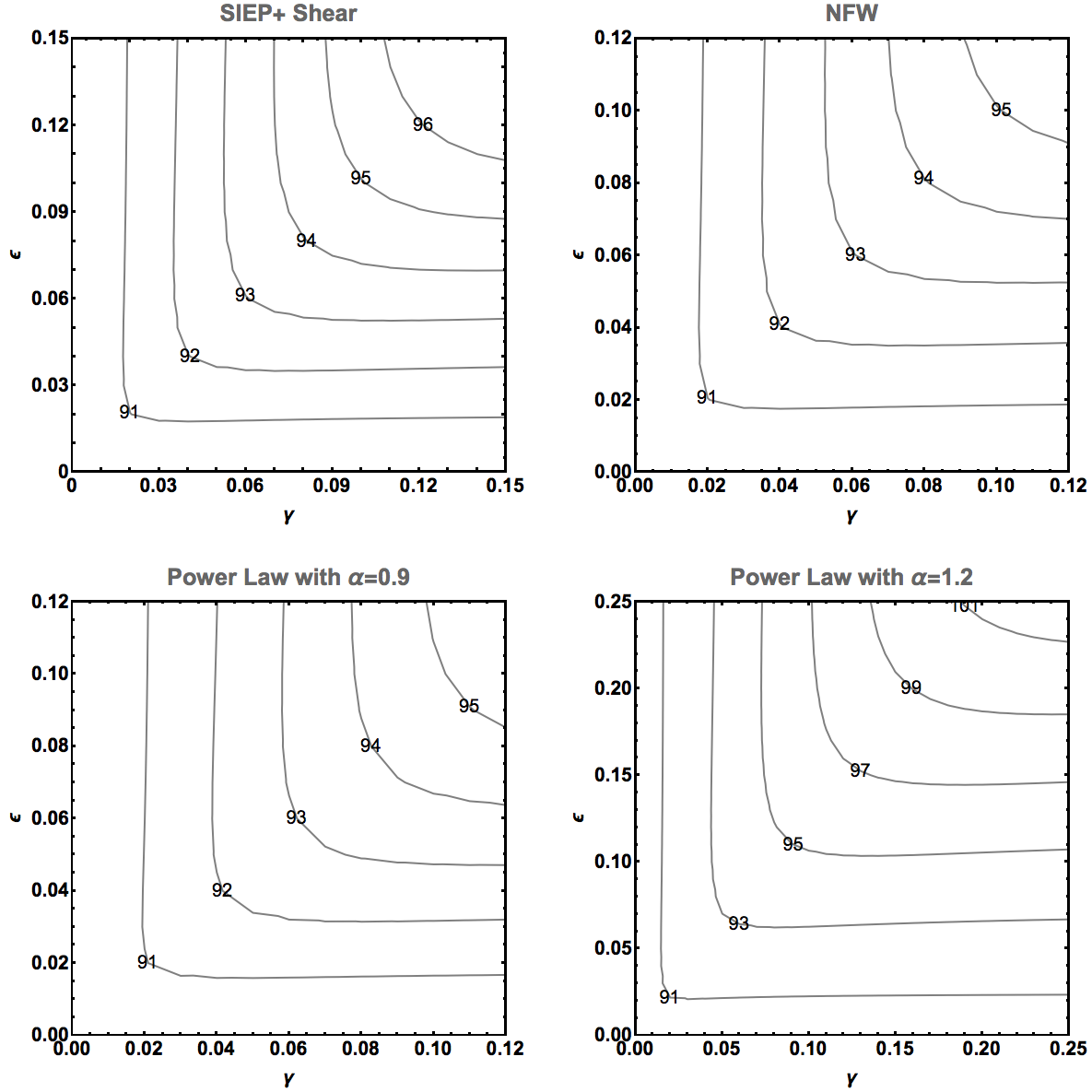}
\caption{Contours of constant $\theta_{23,p}$ in $\gamma$ vs. $\epsilon$ plane for lenses of four different radial density profiles. The contours are labeled by $\theta_{23,p}$. For SIEP+shear (\textit{Top left}) and NFW (\textit{Top right}) lenses the contour lines are symmetric about $\gamma=\epsilon$ line. For hybrid power law potentials the contours are rotated, as compared to that of SIEP+shear, clockwise for $\alpha<1$ (\textit{Bottom left}) and counterclockwise for $\alpha>1$ (\textit{Bottom right}). }
\label{fig6}
\end{flushleft}
\end{figure}

\begin{figure}
\includegraphics[scale=.5]{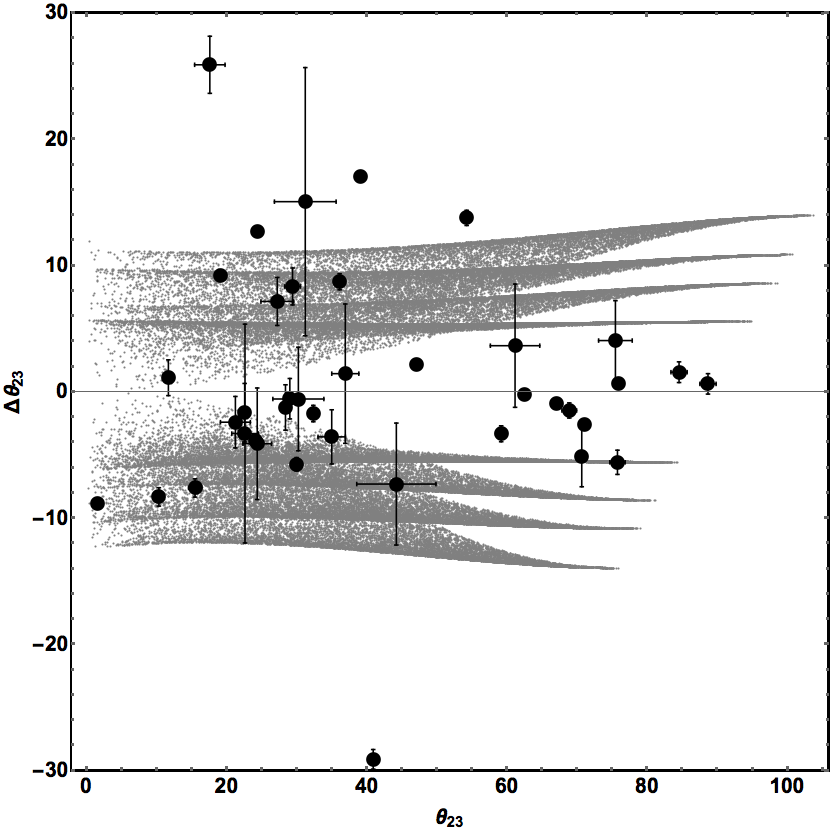}
\caption{Distribution of observed quads (black filled circles) and quads from four Type II lenses (gray distributions) in $\theta_{23}$  vs. $ \Delta\theta_{23}$ plane. The quad data and the errorbars were taken from WW12. $\Delta\theta_{23}=0$ is FSQ. Each lens has two gray surfaces which are about equally close to, but on opposite sides of the FSQ. From the outermost to the innermost surfaces, the lens are: NFW ($\beta=0.1$rad, $\epsilon =0.27, \gamma =0.25$), Power Law ($\alpha=1.2, \beta=0.2$rad, $\epsilon =0.25, \gamma =0.25$), Power Law ($\alpha=1, \beta=0.2$rad, $\epsilon=0.15, \gamma=0.17$), and Power Law ($\alpha=0.9, \beta=0.15$rad, $\epsilon=0.14, \gamma=0.13$), respectively.  }
\label{fig7}
\end{figure}

\begin{figure}
\includegraphics[scale=.37]{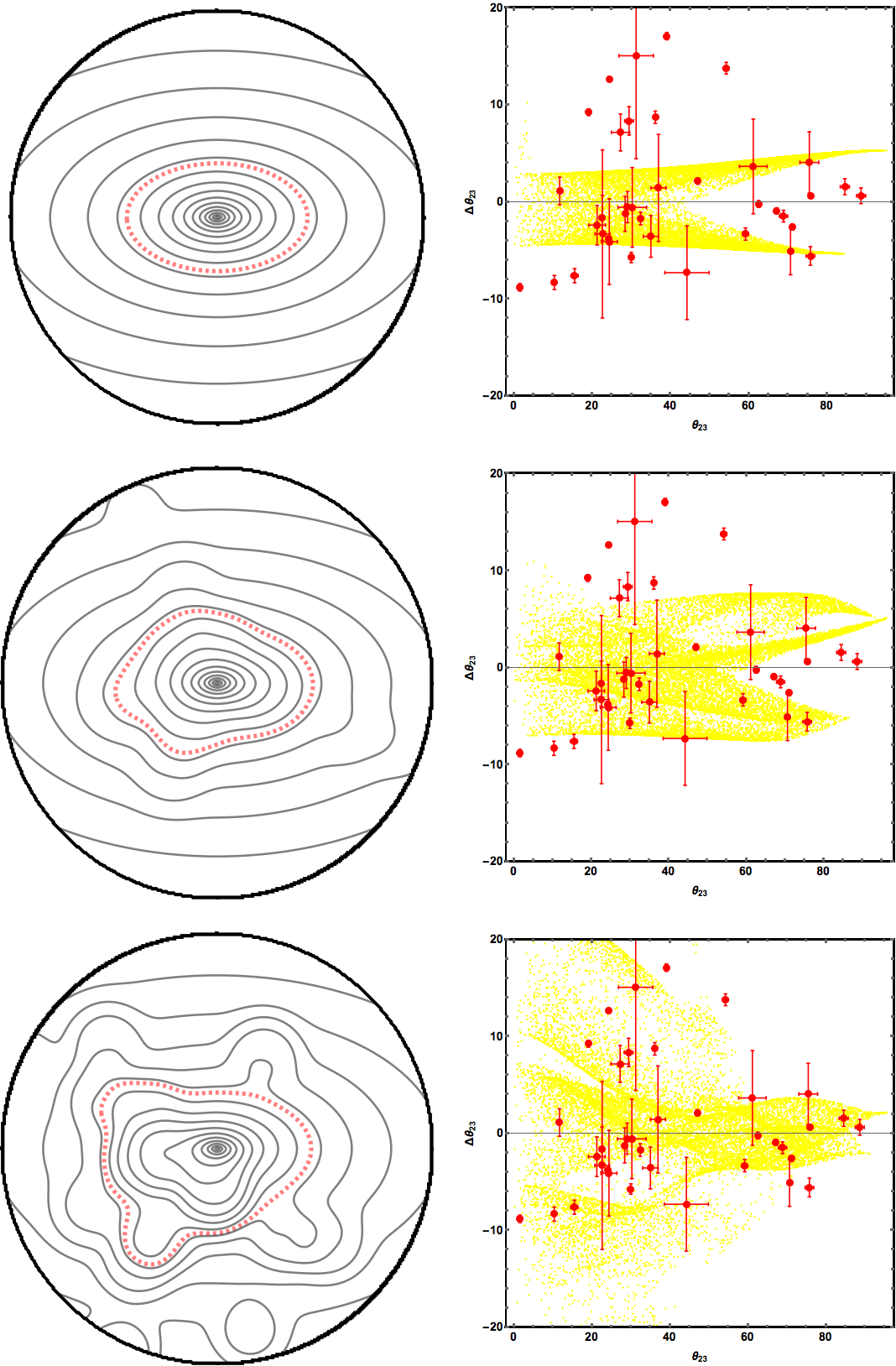}
\caption{\textit{Left panels:} All three have isothermal radial density profile with ellipticity $\epsilon=0.25$ and shear $\gamma=0.2$ oriented at an angle $\beta=\pi/6$ to each other. The $\kappa=1$ contour is shown by the dashed, red, contour. \textit{Right panels:} The corresponding distribution of quads from these lenses in the $\theta_{23}$ vs. $\Delta\theta_{23}$ plane are shown as the the light shade, yellow, distributions. The observed quads (same as in Figure~\ref{fig7}) are red (black) dots.  \textit{Top:} A Type II lens with no substructure. \textit{Middle:} A Type III lens with randomly distributed clumps that account for 4.38\% of the total mass within the window. \textit{Bottom:} A Type III lens with randomly distributed clumps that account for 17.96\% of the total mass within the window.}
\label{fig8}
\end{figure}

\begin{figure}
\includegraphics[scale=.4]{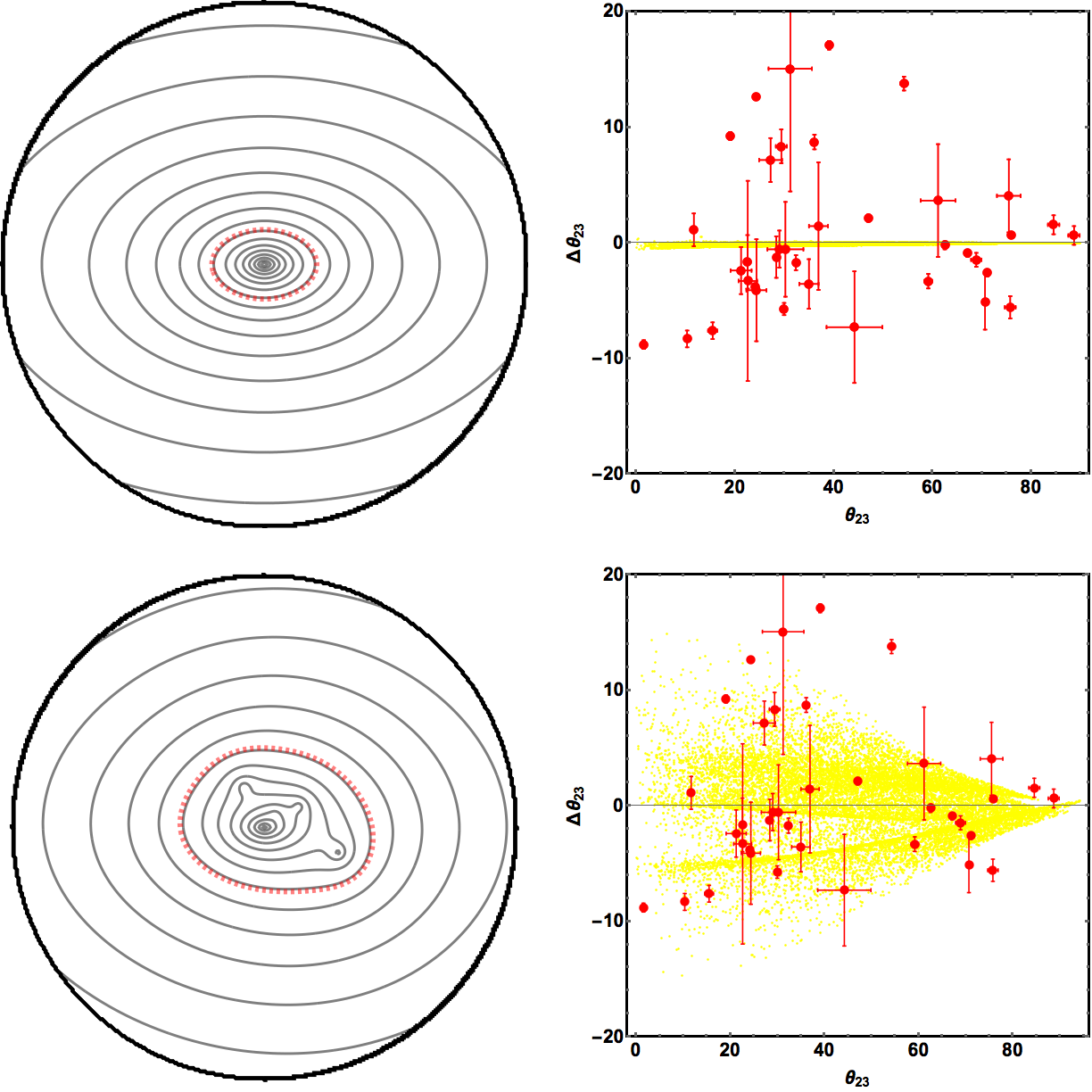}
\caption{Similar to Figure~\ref{fig8}, but different lenses. \textit{Top panels:} A Type I lens with elliptical isothermal radial density profile with $\epsilon=0.25$, and $\gamma=0$. \textit{Bottom panels:} A Type III substructured lens.}
\label{fig9}

\end{figure}

We classify lenses into three main categories based on whether their diamond caustic obey twofold and double mirror symmetries. The caustics of Type I lenses fulfills both symmetries, those of Type II obey only the twofold symmetries while the caustics of Type IIIs obey neither. In this paper we focused on Type II lenses. Our main working space is the 3D space spanned by the three relative image angles of a quad, $\theta_{12}$, $\theta_{23}$, and $\theta_{34}$. One of our main conclusions is that this space is a useful tool in studying quad lenses. We showed how the distribution of the three angles of a given lens relates to its lensing potential and the caustic. The two fold symmetry of the Type II lens caustic is reflected in the fact that in the 3D angle space the distribution of the relative angles is always confined to two surfaces (Figure~\ref{fig2}). This is the defining feature of Type II lenses.

Of the three potentials we studied the closest quantitative connections between the lens potential, caustic and 3D angle space are exhibited by the SIEP+shear potential (eq.~\ref{eq1} with $\alpha=1$). The angle between the second and third arriving images of the source located at the center of the lens, $\theta_{23,p}$ is equal to the angle between the caustic diagonals. (Though we do not study the images distances from lens center in this paper, we note that the distance ratio of these two images is equal to the ratio of the two caustic diagonals.) SIEP+shear lenses that have the same angle $\theta_{23,p}$ have nearly the same distributions of quads in the 3D angle space, implying that $\theta_{23,p}$ is sufficient to specify the shape of the surfaces of all these lenses, and that these surfaces are degenerate. Other lenses also show close similarity between lenses of the same $\theta_{23,p}$, but not as close as for SIEP+shear. The similarity extended to lenses of different radial density profiles: quads of Type II lenses with ellipticities $\epsilon/\gamma\lesssim 2-2.5$ that have the same $\theta_{23,p}$ share approximately the same two surfaces in the 3D angles space.

In addition to the lens potentials discussed in detail in this paper we explored more general forms such as the Sersic profiles. The commonalities in the distribution of quads in the 3D angle space (the general shape and location of the two surfaces) persist in all the models considered. This should not come as a surprise since it is a direct reflection of the two-fold symmetry of the caustics of Type IIs which identify the class. Furthermore, the near degeneracies described above persist across all the models studied and we conjuncture that they extend to all Type IIs since there is no evidence to the contrary. Because these observations apply to a wide range of models and allow us to draw conclusions about Type II lenses as a class, we call this approach model-free analysis. We conclude that, even though Type II lenses can successfully model individual quad lenses, they cannot reproduce the relative image angle distribution of the population of the observed quads. 

We present two examples of Type III  substructured lenses that have distributions of relative image angles that are generally consistent with the observed distribution. These examples are not meant to be a close match to observations. Future work will need to look at different types of substructure, as well as luminous and dark secondary perturber galaxies more carefully to ascertain what type and amount of substructure is necessary to fit the observed quad population. Finally, we note that the substructure mass clumps one would need to explain image angles are likely to be much larger in mass and extent compared to those thought to be responsible for the flux ratio anomalies of quads.

\clearpage{}

\onecolumn

\clearpage

\begin{table}
\begin{center}
 \begin{tabular}{|l|l|l|l|l|}
\hline
Lens 	&  Radial Profile	& L~($\beta,\epsilon,\gamma,\theta_{23,P}$)  &  Cross-validation & Lens comparison\\
name        &                              &                                                                  &   RMSE  (deg.)    &  RMSE (deg.) \\
\hline
A 			& SIEP		  	& L~(1.2, 0.09, 0.10366, 92)                           &  0.005           &  \\
B 			&				& L~(1.2, 0.06603, 0.19, 92)                           &                      & A: 0.011\\
C 			&				& L~(0.5, 0.14, 0.036752, 92)                         &                      & A: 0.029\\
D 			& 				& L~(0.5, 0.034944, 0.06, 92)                         &                      & A: 0.017\\
E 			&				& L~($\frac{\pi}{4}$, 0.069197, 0.08, 93)        &  0.003           & \\

\hline

F			& Power Law $\alpha$=1.2    & L~($\frac{\pi}{6}$, 0.2, 0.0446, 93) & 0.013            & \\
G 			&                                            & L~($\frac{\pi}{6}$, 0.06536, 0.18, 93)&                   & E: 0.146, F: 0.551\\
\cline{2-5}

H 			& Power Law $\alpha$=0.9    & L~($\frac{\pi}{6}$, 0.04706, 0.11, 93)& 0.003         & E: 0.011 \\
I 			&			                       & L~($\frac{\pi}{6}$, 0.095, 0.058, 93) &                    & H: 0.264\\

\hline

J			& NFW	        	& L~($\frac{\pi}{6}$, 0.054, 0.075, 93)                & 0.006          & E: 0.085 \\
JJ			&      	        	& L~($\frac{\pi}{6}$, 0.075, 0.054, 93)                &                    & J: 0.007 \\
K			& 				& L~($\frac{\pi}{6}$, 0.13, 0.05265, 93)              & 0.002          & J: 0.181\\
KK			& 				& L~($\frac{\pi}{6}$, 0.05265, 0.13, 93)              &                   & K: 0.036\\

\hline

\end{tabular}
\end{center}
\caption{Examples of Type II lenses. L~($\beta,\epsilon,\gamma,\theta_{23,P}$) refers to a lens with ellipticity  $\epsilon$, external shear $\gamma$, which is oriented at an angle of $\beta$ (in radians) relative to the principal axis of the main lens, and relative image angle $\theta_{23,p}$ between images 2 and 3 of a central source. The subscript $p$ in $\theta_{23,p}$ indicates that it is the $\theta_{23}$ value of the quad located at the peak of the top surface in the 3D space of relative angles. The cross-validation column lists RMSE values between an interpolation fit to the lens and its own quads. The lens comparison column lists RMSE value between the fit to a lens (listed in column 5) and quads from a different lens (lens named in that row). Lenses JJ and J (and also K and KK) have their $\epsilon$ and $\gamma$ values swapped.}
\label{table1}
\end{table}

\bsp 

\label{lastpage}

\end {document}